\documentclass[twocolumn,aps,pra,showpacs,groupedaddress]{revtex4}

\usepackage{amsmath}
\usepackage{dcolumn}
\usepackage{epsfig}
\usepackage{graphicx}
\usepackage{latexsym}
\usepackage{epstopdf}

\begin{document}

\title{Thermometry and Refrigeration in a Two-Component Mott Insulator of Ultracold Atoms}
\author{David M. Weld}
\email{dweld@mit.edu}
\author{Hirokazu Miyake}
\author{Patrick Medley}
\altaffiliation{Present address: Department of Physics, Stanford University, Stanford CA 94305}
\author{David E. Pritchard}
\author{Wolfgang Ketterle}
\affiliation{MIT-Harvard Center for Ultracold Atoms, Research Laboratory of Electronics, and Department of Physics, Massachusetts Institute of Technology, Cambridge MA 02139}
\begin{abstract}

Interesting spin Hamiltonians can be realized with ultracold atoms in a two-component Mott insulator (2CMI)~\cite{lewenstein-review,zwerger-review}.  It was recently demonstrated that the application of a magnetic field gradient to the 2CMI enables new techniques of thermometry~\cite{thermometrypaper} and adiabatic cooling~\cite{demagarxiv}.  Here we present a theoretical description which provides quantitative analysis of these two new techniques. We show that adiabatic reduction of the field gradient is capable of cooling below the Curie or N\'eel temperature of certain spin ordered phases.

\end{abstract}


\maketitle

The possibility of using ultracold lattice-trapped gases as general simulators of strongly interacting many-body systems has excited increasing interest in recent years~\cite{feynmanQS,lewenstein-review,zwerger-review}.  Spin hamiltonians are a natural candidate for quantum simulation, especially given the relevance of doped antiferromagnetic systems to the important open problem of high-$T_c$ superconductivity~\cite{Lee-highTcdopedMI}.  The 2CMI is the starting point for simulation of electronic spin systems in lattices~\cite{florence-2CMI,bloch2CMIfermions,fermion2CMI,dominik-SFmixtures,sengstock-KRb-2CMI}.  Spin-exchange-stabilized magnetically ordered states are expected to exist in the 2CMI~\cite{duandemlerlukin}, and observation of these states and transitions between them would open up an exciting new field at the intersection of atomic and condensed matter physics.  The main obstacle which has so far prevented the observation of spin-ordered states in the 2CMI is the very low temperature scale required for spin ordering~\cite{ho-mott}.  Quantum Monte Carlo calculations have predicted Curie and N\'eel temperatures on the order of 200 pK for the ferromagnetic and antiferromagnetic states of $^{87}$Rb in a 532-nm lattice~\cite{criticalentropy}. This is a lower temperature than has ever been measured in any system. Clearly, new methods of thermometry and refrigeration are required. 

The recently demonstrated technique of spin gradient thermometry~\cite{thermometrypaper} should allow measurement of temperatures down to the spin exchange scale in the 2CMI.  The related method of spin gradient demagnetization cooling is capable of cooling to the neighborhood of the critical temperature for spin ordering~\cite{demagarxiv}.  Together, these new techniques open a realistic prospect of preparing spin-ordered phases in the 2CMI.  In order to compare experimental results with theory, we have developed a simple theoretical model of the 2CMI and used it to calculate the expected response of our system to spin gradient thermometry and spin gradient demagnetization cooling.  

Our treatment of the 2CMI is similar in approach to the studies of cooling in the one-component Mott insulator presented in Refs.~\cite{ho-mott}~and~\cite{gerbier-mott}, in that it is based on calculations of entropy-versus-temperature curves for various values of control parameters.  Our model neglects the effects of tunneling and treats each lattice site separately, yet is capable of qualitatively reproducing observed cooling curves using only one fit parameter (the initial temperature)~\cite{demagarxiv}.  Our results thus complement, and are in qualitative agreement with, the classical mean field and Monte Carlo analysis of Natu and Mueller~\cite{natu-domainwalldynamics}.

The inputs to the calculation are the measured trap frequencies $\left(\omega_x,\omega_y,\omega_z\right)$, the total atom number $N$, and the applied magnetic field gradient $\nabla|\textbf{B}|$ (along with various fixed parameters like the scattering lengths and magnetic moment of the atoms and the lattice constant).  This allows direct comparison with experiment.  The particular trap frequencies assumed here are $2\pi\times\left(40,156,141\right)${ }Hz.  We assume an atom number of 17000, leading to an occupation number of 3 in the center of the cloud.  These values were chosen because they are typical in our experiments.  The scattering lengths we assumed are $a_{\uparrow\uparrow}=100.4~a_0$,  $a_{\downarrow\downarrow}=98.98~a_0$, and $a_{\uparrow\downarrow}=98.98~a_0$, where $a_0$ is the Bohr radius and states $\uparrow$ and $\downarrow$ are the \mbox{$|F=1,m_F=-1\rangle$} and \mbox{$|F=2,m_F=-2\rangle$} hyperfine states of $^{87}$Rb; these values represent the results of the most recent theoretical calculations available~\cite{kokkelmansprivcomm}.

Detailed technical descriptions of spin gradient thermometry and spin gradient demagnetization cooling are presented in Refs.~\cite{thermometrypaper}~and~\cite{demagarxiv}, respectively.  Both techniques are based on the 2CMI in a magnetic field gradient.  Since the two components have different magnetic moments, the gradient pulls them towards opposite sides of the trap, creating two spin domains which remain in thermal contact.  At zero temperature, there will be a zero-width boundary between the two domains, but at finite temperature a mixed region composed of spin excitations will be present.  

Since the total magnetization is always chosen to be zero, the average value of the magnetic field is cancelled by a Lagrange multiplier and can be subtracted from the real field $\textbf{B}(x)$. This allows us to write the field as $B_{\mathrm{eff}}=\nabla|\textbf{B}|\cdot\mbox{\boldmath$x$}_i$, where $\mbox{\boldmath$x$}_i$ is the vector from the trap center to lattice site $i$ projected along the direction of the magnetic field gradient.  Note that $B_{\mathrm{eff}}=0$ at the trap center.  If tunneling is neglected, then at a magnetic field gradient $\nabla|\textbf{B}|$ the energy of a configuration with $n_\uparrow$ up spins and $n_\downarrow$ down spins at lattice site $i$ is
\begin{eqnarray}
E_i(n_\uparrow,n_\downarrow,\nabla|\textbf{B}|)&=&p\cdot \nabla|\textbf{B}|\cdot x_i \cdot(n_\uparrow-n_\downarrow) \nonumber \\
& & {}+ \frac{1}{2}\sum_\sigma U_{\sigma\sigma} n_\sigma(n_\sigma-1) + U_{\uparrow\downarrow}n_\uparrow n_\downarrow \nonumber \\
& &{}+ V_i\cdot(n_\uparrow+n_\downarrow) - \mu_\uparrow n_\uparrow - \mu_\downarrow n_\downarrow,
\label{energies}
\end{eqnarray}
where $p$ is the amplitude of the effective magnetic moment of the atoms, $x_i=|\mbox{\boldmath$x$}_i|$,  $\sigma = \{\uparrow,\downarrow\}$, $U_{ab}$ is the interaction energy between spin $a$ and spin $b$, $V_i=(m/2)(\omega_x^2x_i^2+\omega_y^2y_i^2+\omega_z^2z_i^2)$ is the optical trapping potential at site $i$, $y_i$ and $z_i$ are the distances of site $i$ from the trap center in the two directions transverse to the gradient, and $\mu_a$ is the chemical potential of spin $a$.  The chemical potential is set by the requirement that the number of atoms of each spin be equal to half the total experimentally measured number.  

\begin{figure}[t]\begin{center}
\includegraphics[width=\columnwidth]{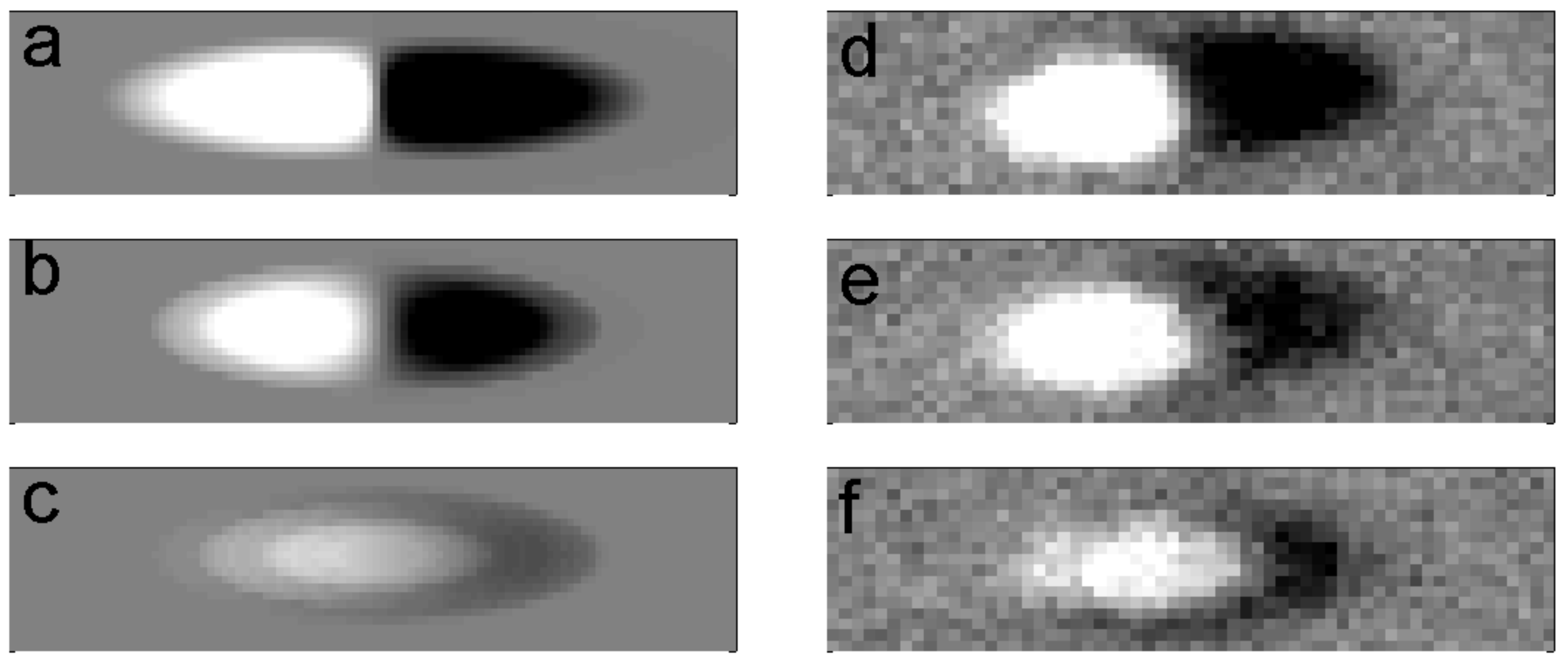}
\caption[demag]{Comparison of simulated and measured spin images.  Simulated images are on the left.  Magnetic field gradients and temperatures for simulated images are: \textbf{a:} 0.7 G/cm, 6 nK, \textbf{b:} 0.06 G/cm, 2 nK, and \textbf{c:} 0.0024 G/cm, 0.4 nK.  The gradient and fitted temperature for each measured spin image \textbf{d-f} are similar to the values for the simulated image in the same row.  See Fig.~\ref{indistinguishability} for a comparison of the temperature extracted from this fit to the real modeled temperature.  Note that the total magnetization in all pictures is close to zero.
\label{simulatedimages}}
\end{center}\end{figure}

\begin{figure}[hbt]\begin{center}
\includegraphics[width=\columnwidth]{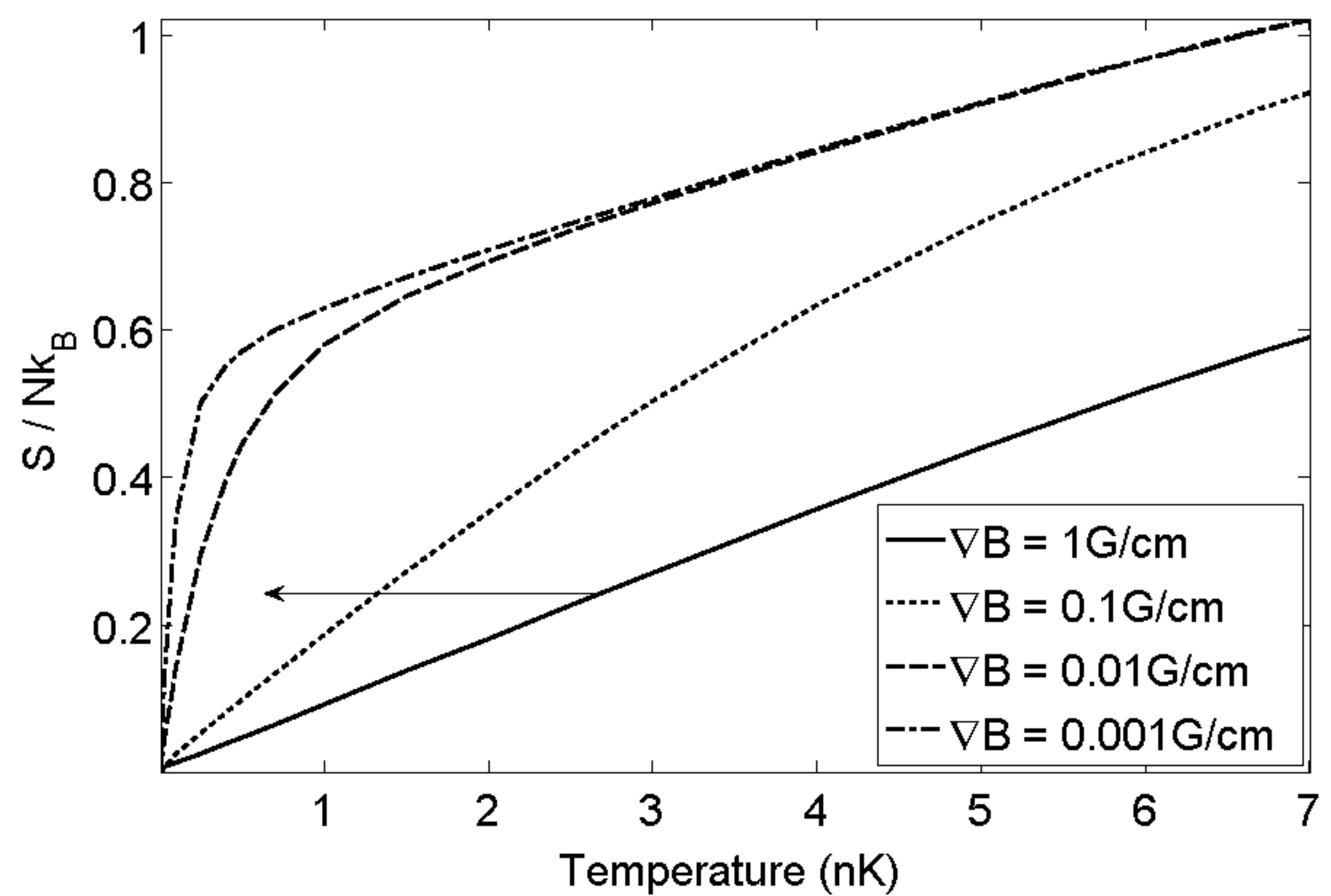}
\caption[demag]{Total entropy per particle versus temperature, at various gradients, for the experimental parameters described in the text.  The arrow indicates a possible path followed during adiabatic spin gradient demagnetization cooling.
\label{SvsT}}
\end{center}\end{figure}

\begin{figure}[hbt]\begin{center}
\includegraphics[width=0.9\columnwidth]{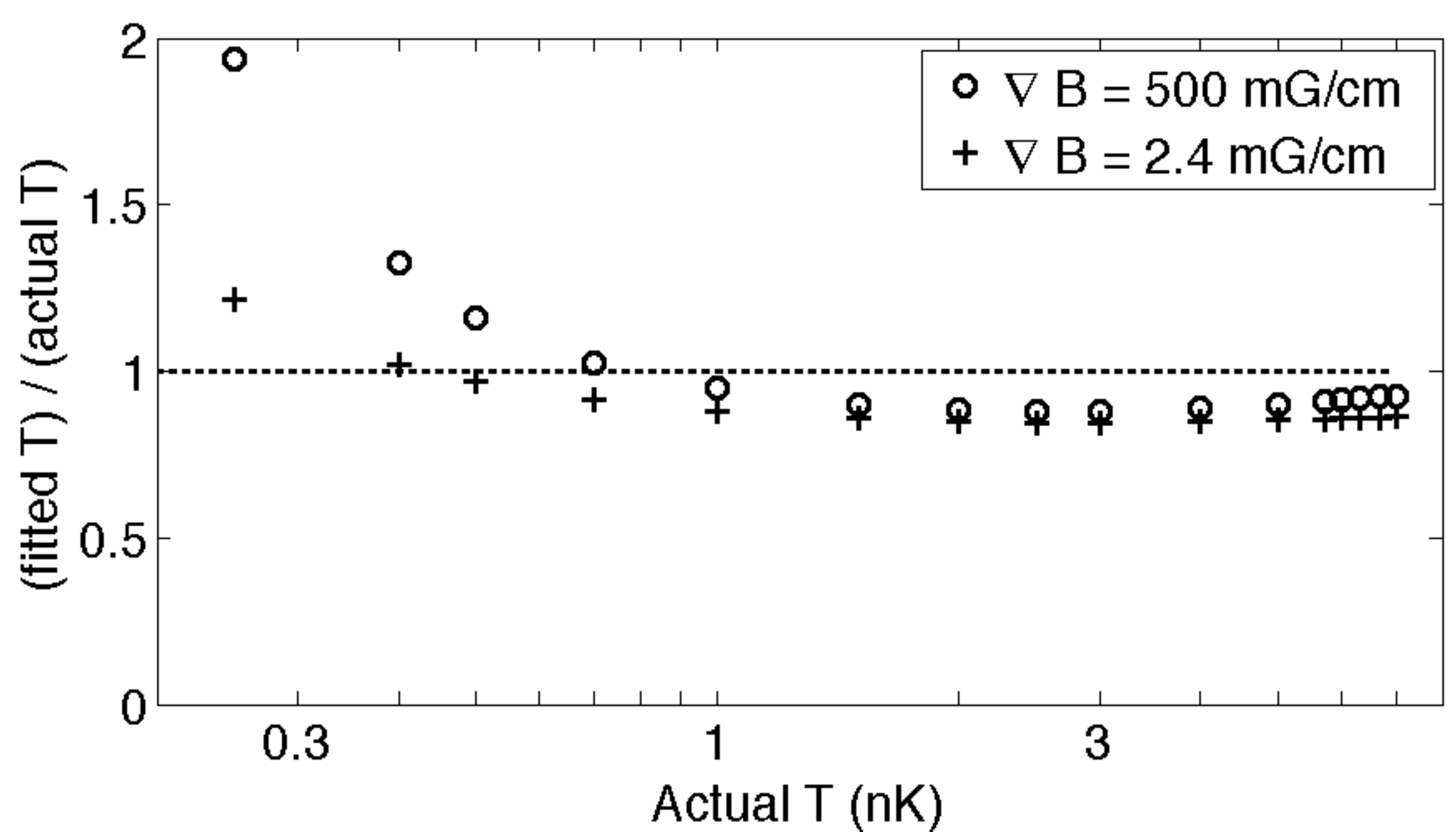}
\caption[demag]{Ratio between fitted temperature and actual temperature at two different gradients, assuming perfect imaging.  This shows the effect of corrections due to indistinguishability and unequal scattering lengths.  Finite imaging resolution will limit the range of temperature that can be measured with any given gradient, as discussed in Ref.~\cite{thermometrypaper}.
\label{indistinguishability}}
\end{center}\end{figure}

\begin{figure}[hbt]\begin{center}
\includegraphics[width=\columnwidth]{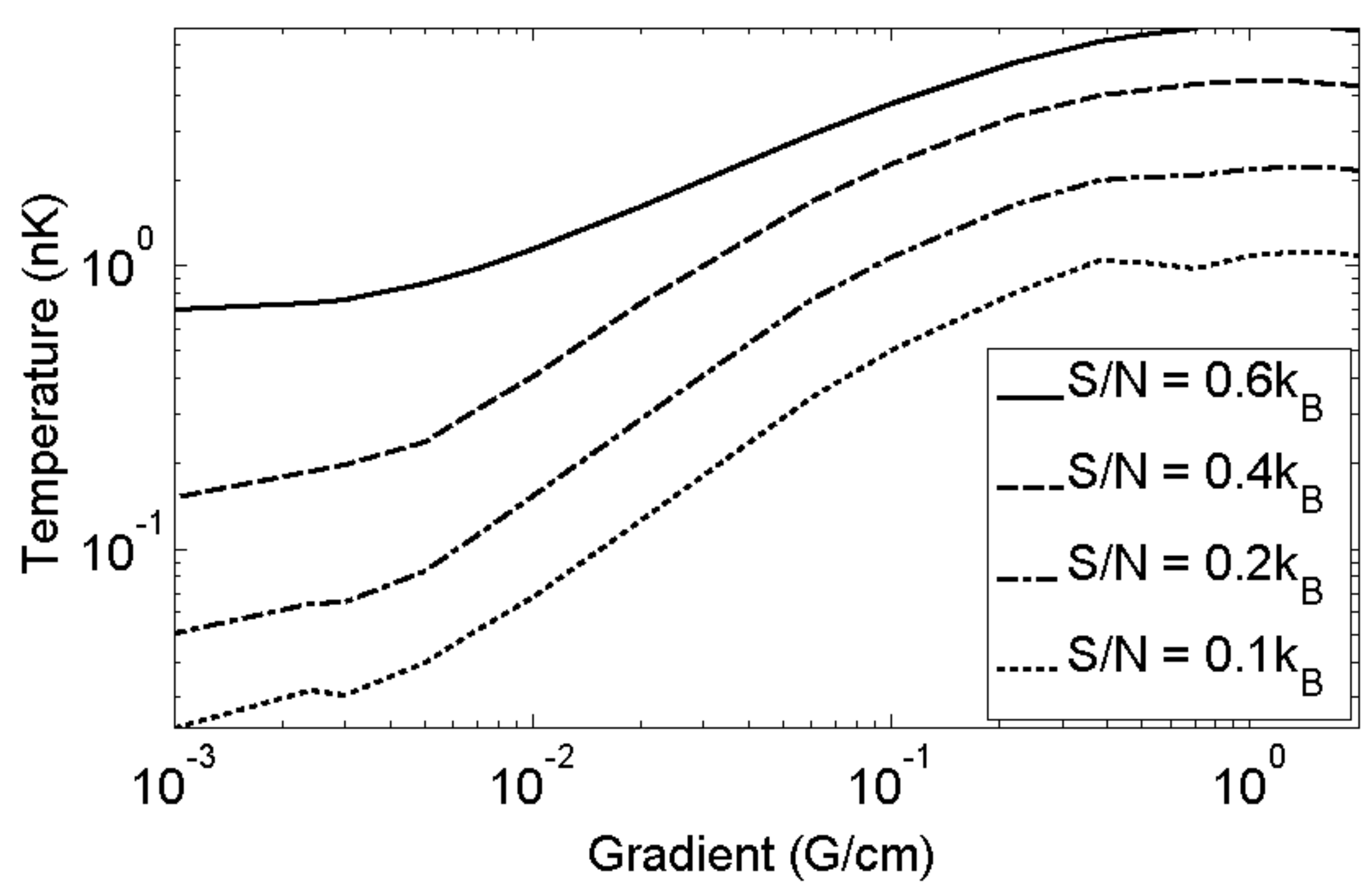}
\caption[demag]{Spin gradient demagnetization cooling.  Predicted temperature versus final gradient, for several values of the total entropy.  
\label{Tvsgrad}}
\end{center}\end{figure}

Equation~\ref{energies} can be used in the grand canonical ensemble to infer the thermal probability of different occupation numbers of the two spins.  The partition function at lattice site $i$ is $Z_i(\nabla|\textbf{B}|)=\sum_{\{n_\uparrow,n_\downarrow\}} \exp(-E_i(n_\uparrow,n_\downarrow,\nabla|\textbf{B}|)/k_B T)$, where $k_B$ is Boltzmann's constant, $T$ is the temperature, and the summation is over all possible combinations of $n_\uparrow$ and $n_\downarrow$ (each combination is counted only once, due to indistinguishability of the atoms).  The probability of having $n_\uparrow$ up spins and $n_\downarrow$ down spins at lattice site $i$ is then
\begin{equation}
p_i(n_\uparrow,n_\downarrow,\nabla|\textbf{B}|,T) = \frac{\exp\left(-E_i(n_\uparrow,n_\downarrow,\nabla|\textbf{B}|)/k_B T\right)}{Z_i}
\end{equation}
and the resulting entropy at site $i$ is 
\begin{equation}
S_i(\nabla|\textbf{B}|,T) = \sum_{\{n_\uparrow,n_\downarrow\}}-p_i \log{p_i}
\label{siteentropy}
\end{equation}
 where the summation is performed in the same way as for the partition function.  The only additional approximation needed is a truncation of the sums over spin configurations. For our experimental parameters, configurations corresponding to a total atom number per site $n$ greater than 4 can be neglected, and we have truncated the sums accordingly.  This truncation is reminiscent of, but more general than, the particle-hole approximation~\cite{ho-mott,gerbier-mott}.   The site entropy $S_i$ of Eq.~\ref{siteentropy} is summed over all lattice sites to extract the total entropy as a function of temperature and field gradient. From this output one can extract column-integrated images (Fig~\ref{simulatedimages}), entropy-versus-temperature curves (Fig.~\ref{SvsT}), and the predicted response to thermometry (Fig.~\ref{indistinguishability}) and cooling (Fig.~\ref{Tvsgrad}). 
 
It is instructive to compare the results of this calculation to those of the simple approximation which treats the spin and particle-hole degrees of freedom separately.  In this approximation, the partition function for an individual lattice site $i$ is assumed to factorize as \mbox{$Z=Z_\sigma Z_0$,} where \mbox{$Z_\sigma= \sum_{\sigma} \exp(-\beta \mbox{\boldmath$p$}_\sigma\!\cdot\! \textbf{B}(x_i))$,} $\beta$ is $1/k_BT$, $\mbox{\boldmath$p$}_\sigma$ is the magnetic moment of the spin $\sigma$, and $Z_0$ is the partition function of the particle-hole degrees of freedom (for which see~\cite{ho-mott,gerbier-mott}).   This simple treatment is valid for the case of one atom per lattice site.  For occupation number $n>1$, there are corrections which are captured by our more complete model.  The first correction arises from a difference $\Delta U$ between the mean of the intra-spin interaction energies $\overline{U_\sigma}$ and the inter-spin interaction energy $U_{\uparrow\downarrow}$.  $\Delta U/\overline{U_\sigma}$ is about $0.007$ for our states. The leading correction changes the magnetic field gradient at the center of the sample $B^{\prime}$ to an effective gradient \mbox{$B^{\prime}(1+(n-1)\Delta U/k_B T)$}.  This becomes important at low temperatures, and destroys the factorizability of the partition function mentioned above.  The second correction is due to indistinguishability of the atoms.  This arises from the quantum mechanical fact that there are three (rather than four) possible spin states for a lattice site with two pseudospin-1/2 atoms.  The size of both corrections is expected to be small, but in order to treat them fully we have developed the more general model described above.   

Under the assumption that \mbox{$Z=Z_\sigma Z_0$,} the mean spin $\langle s \rangle$ as a function of position, field gradient, and temperature has the simple form
\begin{equation}
\langle s \rangle=\tanh(-\beta\cdot\Delta \mbox{\boldmath$p$}\cdot\textbf{B}(x_i)/2),
\label{tanh}
 \end{equation}
 where $\Delta \mbox{\boldmath$p$}$ is the difference between the magnetic moments of the two states.  Spin gradient thermometry is based on the fact that at finite temperatures, the width of the boundary layer is proportional to the temperature and inversely proportional to the magnetic field gradient.  
  
  \begin{figure}[h]\begin{center}
\includegraphics[width=1\columnwidth]{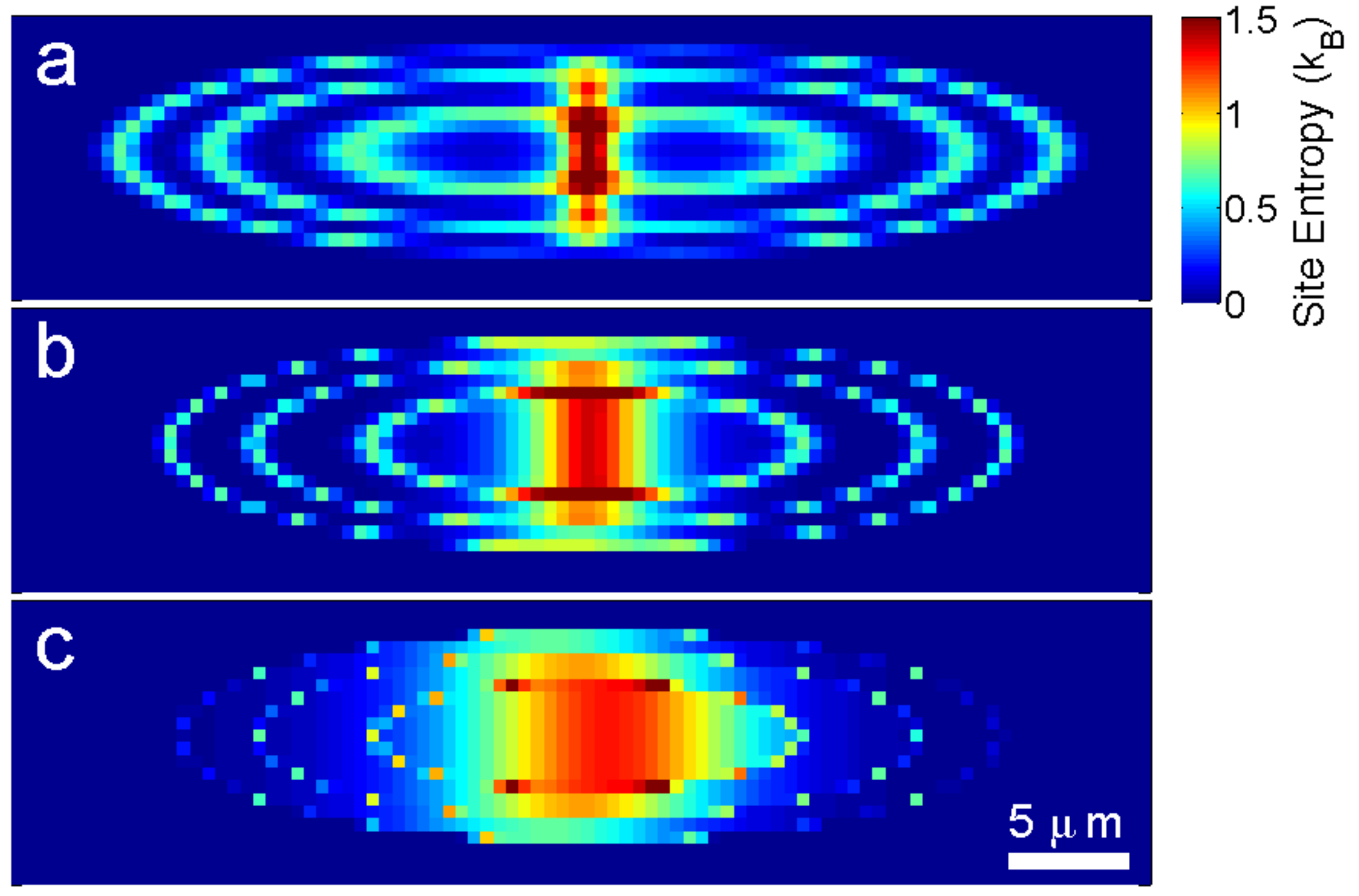}
\caption[demag]{Entropy distribution during spin gradient demagnetization cooling.  The images are slices of the cloud through the center.  Each pixel represents one lattice site. All plots are at a total entropy per particle near 0.3 $k_B$, and are thus representative of the changing entropy distribution during isentropic demagnetization.  Values of the gradient and temperature are as follows: \textbf{a:} $\nabla B=0.5\mathrm{G/cm}, T=3\mathrm{nK}$, \textbf{b:} $\nabla B=0.1\mathrm{G/cm}, T=1.5\mathrm{nK}$, \textbf{c:} $\nabla B=0.02\mathrm{G/cm}, T=0.5\mathrm{nK}$.  The ring-shaped structures are the mixed-occupation-number regions between Mott domains which carry particle-hole entropy.  Note that these regions are narrower after reduction of the gradient.  This indicates that entropy has been transferred from the mixed-occupation-number regions to the mixed-spin region, and the temperature has been reduced.
\label{demagsnapshots}}
\end{center}\end{figure}
  
The spin profile of Eq.~\ref{tanh} is exact for a 2CMI with one particle per site.  The model presented here can be used to investigate corrections to thermometry at higher filling.  Figure~\ref{indistinguishability} shows the temperature measured by fitting to Eq.~\ref{tanh} divided by the actual modeled temperature for two values of the gradient.  The high-temperature correction is mainly due to indistinguishability of the atoms, and is only important for sites containing 2 or more atoms.  Note that the fitted spin profile is integrated along both directions transverse to the gradient, so it includes contributions from all occupation number domains.  Though this correction is conceptually important, it changes the measured temperature by less than 15\% under our experimental conditions.  The correction at low temperatures is partly due to the fact that the scattering lengths $U_{\uparrow\uparrow}$, $U_{\downarrow\downarrow}$, and $U_{\uparrow\downarrow}$ are not all equal.  This is expected to result in a curvature of the mixed region between the two spin domains.  This curvature arises from a buoyancy effect-- the species with greater intraspin repulsion will preferentially populate the outer regions of the trap.  This effect causes a fit to Eq.~\ref{tanh} to overestimate the temperature, since a curved boundary appears wider after integration along the directions perpendicular to the gradient.  Another correction at low temperatures arises if the width of the mixed region becomes much less than one lattice constant.  In this case both the model and the real physical spin system will not respond measurably to small changes in the gradient, and the measured temperature will overestimate the real temperature.  These corrections need to be taken into account for precision temperature measurements at extreme temperatures and field gradients, but they do not alter the conclusions of Refs.~\cite{thermometrypaper}~or~\cite{demagarxiv}.  

Spin gradient demagnetization cooling is based on the fact that the entropy stored in the mixed region between the two spin domains increases with decreasing magnetic field gradient.  If the change of the gradient is adiabatic, then the energy and entropy which flow to the spin degrees of freedom must come from other degrees of freedom, and the sample's temperature can be reduced.  Although spin gradient demagnetization cooling was inspired by (and is locally similar to) adiabatic demagnetization refrigeration in condensed or gaseous systems~\cite{giauque-ADR,debyeADR,pfaudemag}, there are important differences between the techniques.  Most notably, spin gradient demagnetization cooling varies a magnetic field \emph{gradient} rather than a homogeneous field, and relies on spin transport rather than spin flips.  These differences allow the technique to be applied to lattice-trapped ultracold atomic systems.  Spin gradient demagnetization cooling thus broadens and extends existing magnetic refrigeration techniques.  

Entropy versus temperature curves such as those plotted in Fig.~\ref{SvsT} can be used to calculate the response of the system to spin gradient demagnetization cooling.  If the gradient is reduced perfectly adiabatically, the system will move horizontally as indicated by the arrow in Fig.~\ref{SvsT}, and the temperature will decrease.  This behavior is plotted in Fig.~\ref{Tvsgrad} for several values of the total entropy (corresponding to different initial temperatures).  These predictions can be used directly to fit experimental data, with the initial temperature being the only free parameter.  Such fitting gives reasonable agreement (see Ref.~\cite{demagarxiv}).  

For sufficiently low initial entropy, the spin degrees of freedom will contain all the entropy in the system when the gradient is adiabatically reduced by some factor.  Further reduction of the gradient below this point is expected to linearly decrease the temperature of the system until the point where interactions become important.   Conversely, if the initial entropy is too high, the spins will become fully disordered at some finite value of the gradient and will no longer be able to absorb entropy.  Reduction of the gradient below this point will not change the temperature.  This behavior, which is essentially a finite-size effect, is apparent in the upper curve in Fig.~\ref{Tvsgrad}.  If the gradient is sufficiently high, it can pull the two spin domains so far apart that the area where they overlap is decreased in size.  This effect reduces the entropy capacity of the spin degrees of freedom at high gradients, and is the origin of the slight downturn in temperatures at the highest gradients in Fig.~\ref{Tvsgrad}.

Magnetic field gradients of 1 mG/cm are well within the range of the experimentally achievable.  Assuming reduction of the gradient from 2 G/cm to 1 mG/cm, our analysis predicts that samples with an initial entropy lower than about 0.4$k_B$ can be cooled below the spin ordering temperature.  Our model neglects spin correlations, so the lowest-temperature results plotted in Fig.~\ref{Tvsgrad} should be taken as evidence that reduction of the gradient is capable of cooling below the spin ordering temperature rather than a prediction of the dependence of temperature on gradient below the Curie or N\`eel temperature.

Figure~\ref{demagsnapshots} shows several images of the total entropy distribution at different final gradients during demagnetization.   The pumping of entropy from the kinetic degrees of freedom to the spins is clearly visible.

These theoretical results help elucidate some limitations on and possible extensions to the technique of spin gradient demagnetization cooling.  The technique clearly requires the use of two states with different magnetic moments-- this excludes, for example, the two lowest states of $^7$Li at very high fields.  The predicted behavior is also in principle different for higher filling factors than it is for $n=1$ (although as discussed above we find this effect to be small for the particular case of $^{87}$Rb).  For example, strong miscibility or immiscibility of the two species would change the response of the system to demagnetization, but only if the maximum occupation number $n$ is greater than 1 (see also Ref.~\cite{hofstetterlatticephasediagram}).  The dependence of the response to demagnetization on the trap frequencies and total atom number can also be investigated using the techniques presented here; the most important effect of varying these parameters is generally to change the maximum occupation number and the spectrum of particle-hole excitations.  For best cooling performance, the initial entropy should be lower than the \emph{maximum} mixing entropy ($k_B \log 2$ per site for the $n=1$ case).  We believe that spin gradient demagnetization cooling could in principle be applied to fermionic mixtures as well.  In fact, the technique does not even require a lattice, and could potentially be applied to the thermal fraction of a trapped two-component gas.

We have presented results of calculations based on a theoretical model of the 2CMI and its response to a varying magnetic field gradient.  Our results provide quantitative support for spin gradient thermometry and spin gradient demagnetization cooling.

We thank Eugene Demler, Takuya Kitagawa, David Pekker, and Servaas Kokkelmans for useful and interesting discussions.  H.M. acknowledges support from the NDSEG fellowship program.  This work was supported by the NSF, through a MURI program, and under ARO Grant No. W911NF-07-1-0493 with funds from the DARPA OLE program.  Correspondence and requests for materials should be addressed to D.M.W.~(email: dweld@mit.edu).

\end{document}